\documentclass[paper,revised]{geophysics}

\usepackage{xcolor} % Include the xcolor package
\usepackage{amsmath}
\usepackage{amssymb}
\usepackage{booktabs}
\usepackage{siunitx}

\begin{document}

\title{Are Fourier Neural Operators Really Faster for Time-Domain Wave Propagation?}

\renewcommand{\thefootnote}{\fnsymbol{footnote}} 

\ms{Are Fourier Neural Operators Faster} % paper number

\address{
\footnotemark[1]Avathon, \\
12708 Riata Vista Cir Suite B-100, \\
Austin, TX 78727 \\}
\author{Dimitri Voytan\footnotemark[1] and Litan Li\footnotemark[1]}

\footer{Are Fourier Neural Operators Faster}
\lefthead{Voytan \& Li}
\righthead{Are Fourier Neural Operators Faster}

\maketitle

\begin{abstract}
Fourier Neural Operators (FNOs) have been promoted as fast, mesh-invariant surrogates for partial-differential equation solvers, with seismic studies reporting orders-of-magnitude speedup over classical methods. We revisit those claims by benchmarking a state-of-the-art Tucker-tensorized FNO (TFNO) against highly optimized, GPU-accelerated finite-difference (FD) codes for four wave-physics formulations: isotropic acoustic, TTI-acoustic, isotropic elastic, and VTI-elastic. To isolate inference cost, we do not train networks as the runtime depends on architecture, and not specific weight values. When TFNO and FD share the same spatial grid (\(10\;\mathrm{m}\) cells) but TFNO runs at the temporal Nyquist rate (coarser than the CFL-limited FD step), TFNO is \emph{consistently slower}, by \(1.4\times\)–\(80\times\), across all physics. Allowing TFNO to operate on a Nyquist-rate spatial grid (\(2.5^3\) coarser than FD) narrows the gap: small architectures (4–8 layers, 12–24 hidden channels) achieve up to \(10\times\) \emph{faster} runtime (VTI-elastic), whereas larger networks (12 layers, 36 channels) produce mixed results: Elastic cases are \(1.09\times\) and \(1.30\times\) \emph{faster} (isotropic and VTI, respectively), while TTI-acoustic and isotropic acoustic are \(1.41\times\) and \(5.9\times\) \emph{slower}. A decisive advantage appears only in a “surface-only’’ configuration that predicts the boundary time history \(u(x,y,z{=}0,t)\): TFNO can be up to \(4\times10^{3}\) \emph{faster} than FD. Overall, TFNOs offer architecture-dependent gains and, for full-volume high-fidelity simulations at matched resolution, are generally slower than optimized FD; by contrast, for boundary-field predictions they can deliver orders-of-magnitude acceleration.
\end{abstract}

\section{Introduction}

The numerical simulation of partial differential equations (PDEs) is a significant area of interest in exploration geophysics, and the development of rapid and accurate solution methodologies has long been a focus of research \cite[]{smith1985numerical, virieux1986p, komatitsch1999introduction}. Recently, surrogate methods based on deep neural operators were introduced as an alternative approach to solving general PDEs \cite[]{kovachki2023neural}, with specific application to seismic wave propagation problems e.g. \cite[]{yang2021seismic, song2022high, zhang2023learning, lehmann20243d}. In general, neural operators are designed to learn mappings between infinite-dimensional function spaces. For problems in seismic wave propagation, salient mappings are from velocity models and wavelets to pressures and/or displacements. The key purported advantages of neural operators include discretization invariance and a fast simulation time---relative to classical methods---for solutions associated with new sets of parameters.

Prior work in applying neural operators to seismic wave propagation problems has focused primarily on training and generalization challenges rather than detailed benchmarking of performance profiles. Applications include forward modeling in both the time and frequency domains, differentiable surrogates for inversion, and physics-aware architectures.

\cite{yang2021seismic} pioneered early work, applying Fourier Neural Operators to learn solutions to the 2D time-domain acoustic wave equation. They train a  network with pairs of randomly generated velocity models based on the von Karman covariance function and associated numerically prepared wavefields. Using the trained model they perform full-waveform inversion via automatic differentiation of the network outputs with respect to the inputs. They report that the FNO-based inversion is at least an order of magnitude faster than an inversion based on a spectral-element code. 

\cite{song2022high} study the use of Fourier Neural Operators to extrapolate high-frequency wave simulations. They work with the 2D frequency-domain acoustic wave equation and show that the Fourier Neural Operator can extrapolate low-frequency wave simulations to higher frequency. They report that the FNO is almost two orders of magnitude faster than a finite difference simulation.

\cite{zhang2023learning} investigate time and frequency domain solutions to the 2D elastic wave equation. They propose a one-connection FNO, which calculates derivatives in multiple directions and adds connections between Fourier layers. They report that ``the modified FNO operates at approximately 100 times the speed of traditional finite-difference methods on a CPU."

\cite{lehmann20243d} train a model to learn a solution operator for the 3D isotropic elastic wave equation. They restrict the operator to learn only the surface wave field, $\mathbf{u}(x,y,z=0)$. This restriction leads to more efficient training and inference, but potentially limits the viability of the method to solutions on the same surface, or requires continuation of wave fields to such a surface. They report a mean speedup factor on the order of $10^7$ between the Fourier Neural Operator run on a single NVIDIA-100 GPU and their spectral element code, run on a cluster of CPUs.

In this paper, we re-examine the claim that Fourier Neural Operators (FNOs) are faster than classical solvers for time-domain seismic modeling. We benchmark a state-of-the-art Tucker-tensorized FNO (TFNO) against GPU-accelerated finite-difference (FD) codes across four canonical formulations: isotropic acoustic, TTI-acoustic, isotropic elastic and VTI-elastic. When both solvers share the same $10\times10\times10$m grid, the TFNO is consistently slower, by factors ranging from 1.4 to more than $80 \times$. At the spatial Nyquist grid (\(25\times25\times25\;\mathrm{m}\)), clear speed-ups occur only for small networks: with 4 layers and 12 hidden channels we measure up to \(9.65\times\) (VTI-elastic; \(8.07\times\) elastic; \(5.31\times\) TTI-acoustic; \(1.25\times\) acoustic), whereas larger networks, 36 hidden channels, 12 layers provide at most \(1.30\times\) (VTI-elastic; \(1.09\times\) elastic) with slowdowns elsewhere. In all experiments, we assume a temporal Nyquist rate for TFNO, while finite difference simulations are constrained by the CFL limit \cite[]{courant1967partial}. A decisive speed-up (up to $4 \times 10^3$) appears solely in a ``surface-only" configuration, as proposed by \cite{lehmann20243d}, where the network is tasked to predict a full time history on the $z=0$ domain surface, i.e. $u(x,y,z=0,t)$. Together with simple asymptotic analysis, our measurements indicate that, on identical hardware and comparable grids, FNOs do not deliver the large speedups often reported. Competitive cost typically requires Nyquist-rate coarsening and small-capacity networks. The surface-only case is more favorable but changes the task and introduces practical hurdles.

Emphatically, we do not focus on the network training time or the challenges associated with learning a solution operator---such as out-of-distribution generalization---explored in previous work. For the purpose of comparing the network inference cost, \textit{we do not train any models}. This simplification is justified because the computational complexity of inference in neural networks is determined by the architecture (for FNO in particular the number of layers and hidden channel dimensions) rather than the specific weight values learned during training. The floating-point operations required for forward propagation through the network remain constant regardless of the trained parameters, making it possible to analyze inference costs independently of training considerations.

In our benchmarking, we emphasize fair and consistent comparisons by carefully controlling for key factors that can significantly influence runtime. All benchmarks are conducted on the same hardware platform to eliminate discrepancies arising from differences in computational architecture, especially between CPU and GPU implementations. Code execution is optimized using appropriate compiler flags and low-level performance tuning where applicable, ensuring that both neural network and finite-difference solvers operate at or near peak efficiency. We also account for the fact that frameworks such as PyTorch employ highly optimized CUDA kernels, which may not offer a fair baseline when compared against naïvely implemented solvers. Therefore, when comparing methods, we aim to match the level of optimization and hardware utilization to provide an accurate picture of relative computational costs. This ensures that the performance characteristics of different approaches can be meaningfully compared in terms of their scalability and practical utility.

The paper is organized as follows. We begin with a brief review of finite difference methods for wave propagation, focusing first on the theoretical considerations related to cost. We then discuss how we use Devito \cite[]{luporini2020architecture} to prepare efficient computational implementations. Next, we introduce Fourier Neural Operators, focusing on theoretical foundations and recent advances. We then compare the asymptotic cost of the Fourier Neural Operator forward pass to finite-difference simulations. We describe our computational benchmarking experiments and present the results of these experiments. Finally, we discuss the implications of our findings and areas of future work. 

\subsection{Background}

% \textcolor{blue}{Literature review}

\subsection{Finite Difference Methods for Wave Propagation and Computational Cost}

Consider the 3-D constant–density acoustic wave equation

\begin{equation}
    \frac{\partial^2 u}{\partial t^2} = c^2(x,y,z) \left( \frac{\partial^2 u}{\partial x^2} + \frac{\partial^2 u}{\partial y^2} + \frac{\partial^2 u}{\partial z^2} \right),
    \label{eq:wave_equation}
\end{equation}
with suitable initial and boundary conditions, and where \( u = u(x, y, z, t) \) is the acoustic pressure, \( c \) is the wavespeed in the medium, and \( t \) is time.

Discretization of space on a uniform grid of $N = nx \times ny \times nz$ nodes with spatial indices $i, j, k$, temporal index, $n$, and with time-step size \( \Delta t \), a $p$-point stencil for the Laplacian leads to the explicit update

\begin{equation}
\label{eq:time_stepping_general}
u^{n+1}_{i,j,k}=2u^{n}_{i,j,k}-u^{n-1}_{i,j,k}
+(\Delta t)^{2}\,c^{2}_{i,j,k}\;\mathcal \nabla^2_{p}u^{n}_{i,j,k},
\qquad
\end{equation} 
\begin{equation}
\mathcal \nabla^2_{p}=\sum_{|\ell|\le (p-1)/2}\!w_{|\ell|}\,u_{i+\ell_{x},j+\ell_{y},k+\ell_{z}}
\end{equation}
where the weights \(w_{|\ell|}\) are chosen to match the
continuous Laplacian to \(p^{\text{th}}\)-order accuracy. Each update touches
\(\mathcal{O}(p^{d})\) neighbors per node, giving a per-step cost
\(\mathcal{O}(p^{d}N)\) and total cost
\(\mathcal{O}(p^{d}N\,T)\) for \(T\) steps.

To control numerical dispersion the grid spacing must resolve
the shortest wavelength \(\lambda_{\min}=c_{\min}/f_{\max}\) by at
least \(\beta\) points with \(\beta\approx10\) for a 2nd order
stencil, and smaller for higher order \cite[]{alford1974accuracy, levander1988fourth, liu2013time}. Hence
\(N\propto\beta^{d}\) for fixed physical volume, while the CFL
condition \(C=c\,\Delta t/\Delta x\le 1\) couples the temporal
and spatial refinements \cite[]{courant1967partial}. In practice, slow wavespeeds and high-frequency simulations increase $N$ leading to a corresponding temporal refinement to satisfy CFL and exponential growth in cost.

The scheme is easily parallelized, as each grid point is
updated from its local stencil, leading to efficient hardware implementation.  These asymptotic scalings define the finite-difference baseline against which
we compare the neural operator performance.

\subsection{Finite Difference Computation with Devito}

We use Devito to implement the wave propagation codes presented herein. Devito is a Python package designed to generate optimized and compiled C/C++ finite difference code from a symbolically described PDE \cite[]{luporini2020architecture}. It is particularly well-suited for explicit time-marching schemes, offering a high level of performance optimization `out-of-the-box'. The package supports various back-end accelerations, including GPU acceleration via OpenACC. 

For CPU simulation, Devito makes several advanced performance optimizations during code generation. These optimizations are implemented through multiple compilation passes, each targeting specific aspects of computational efficiency. Several key optimizations are applied during code generation:

\begin{enumerate}
    \item Loop Blocking: Restructures loops to work on smaller blocks of data that fit into cache memory, optimizing cache usage, and reducing memory access latency. Block sizes are determined heuristically but can be further tuned.
    \item Code Motion: Lifts time-invariant sub-expressions out of inner loops, reducing redundant calculations.
    \item Flop-Reducing Transformations: Includes techniques such as Common Sub-expression Elimination (CSE), factorization, and the replacement of power operations with multiplications, all aimed at reducing the number of floating-point operations.
    \item Cross-Iteration Redundancy Elimination: Identifies and removes redundant computations across loop iterations.
\end{enumerate}
These optimizations collectively offer significant performance improvements, even with default settings. With additional tuning, users can achieve even greater efficiency.

For GPU-based computations, these CPU-oriented optimizations are often unnecessary due to the fundamentally different architecture of GPUs. With their massively parallel structure and high memory bandwidth, GPUs require distinct optimization strategies. These focus primarily on maximizing parallelism, coalescing memory accesses, and efficient shared memory utilization. The open-source version of Devito uses OpenACC pragmas to facilitate these optimizations, providing a high-level abstraction for parallel programming. This enables the compiler to generate code that efficiently maps computations onto the GPU's architecture. 

\subsection{Neural Operators}

Here, we first review operator learning theory as presented by \cite{li2020fourier}. Then, we introduce the FNO architecture. Finally, we detail the TFNO architecture, detailing potential sources of computational cost. 

Let $\mathcal{A} := {a: D \to \mathbb{R}^{d_A}}$ and $\mathcal{U} := {u: D \to \mathbb{R}^{d_U}}$ denote input and output function spaces, respectively. Here, $D \subset \mathbb{R}^d$ is a bounded open set representing the input domain and $d_A, d_U \in \mathbb{N}$ are the dimensions of the input and output spaces. A neural operator is trained to learn the mapping $\mathcal{G}: \mathcal{A} \rightarrow \mathcal{U}$ given a dataset ${(a_j, u_j)}_{j=1}^N$ where $G(a_j) = u_j$. Two classes of operator networks have recently emerged to approach this task, namely DeepONets \cite[]{lu2021learning}, and Fourier Neural Operators \cite[]{li2020fourier, azizzadenesheli2024neural}.

DeepONets are a type of neural network architecture based on the universal approximation theorem for operators \cite[]{chen1995universal}. The network is split into a trunk and a branch component, based on a finite approximation of this theorem. Since we do not focus on this work herein, we refer interested readers to \cite{lu2021learning}. Fourier neural operators, on the other hand, parameterize the operator using a composition of layers $v_j$ for $j = 0, 1, ... T-1$. The inputs $a \in \mathcal{A}$ are elevated to a higher-dimensional representation by an operator $v_0(x) = P(a(x))$ (usually a linear transformation). Updates continue in the higher-dimensional representation with the scheme:

\begin{equation}
v_{t+1} := \sigma\bigg(Wv_t(x) +  \big(\mathcal{K}(a;\phi)v_t\big)(x)\bigg), \ \ \ \ \ \  \forall x \in D, 
\end{equation}
where $\sigma$ is a (typically nonlinear) activation function, $W$ is a linear transform, and 

\begin{equation}
\big(\mathcal{K}(a; \phi)v_t\big)(x) := \int_D \kappa(x, y, a(x), a(y); \phi)v_t(y) \ dy, \ \ \ \ \ \  \forall x \in D
\label{eq:integral_kernel}
\end{equation}

To implement this integral kernel, \cite{li2020fourier} remove the dependence of $a$ from \ref{eq:integral_kernel} and impose $\kappa_\phi(x, y) = \kappa_\phi(x-y)$, which is a convolution operator. Under the convolution theorem, this can be represented in the Fourier domain as

\begin{equation}
\label{eq:fourier_integral_kernel}
\big(\mathcal{K}(a; \phi)v_t\big)(x) = \mathcal{F}^{-1}\big(\mathcal{F}(\kappa_\phi) \cdot \mathcal{F}(v_t)\big)(x), \ \ \ \ \ \  \forall x \in D
\end{equation}

In the neural network, the term $\mathcal{F}(\kappa_\phi)$ is parameterized as a truncated Fourier series up to $k_{\textrm{max}}$ modes via a complex-valued tensor $R_\phi$. So long as the domain is uniformly discretized, the Fourier transforms can be implemented with $\mathcal{O}(N\log(N))$ computational efficiency via the fast Fourier transform. 

Subsequent advances have proposed improvements to this architecture. For example, \cite{kossaifi2023multi} propose a Fourier neural operator with multi-grid input decomposition (MG-FNO) and low-rank tensor factorization of the network weights (TFNO), collectively (MG-TFNO). The TFNO improves the memory efficiency of FNO by introducing a low-rank decomposition (specifically a Tucker decomposition) \cite[]{tucker1966some}, a type of higher-order singular value decomposition. This effectively places a low-rank constraint on the model parameters.

We use the TFNO architecture in the presented work, so we detail it here. 
In the TFNO, the Fourier weights of every spectral-convolution layer are gathered into a tensor $(T\in \mathbb{C}^{\alpha^{d}\times n\times m})$, where the first \(d\) indices enumerate the retained low-wavenumber modes ($\alpha$ in each spatial direction) and the last two indices map input to output channels.  When a network contains \(L\) such layers with identical channel dimensions (\(m=n\)), the tensors \(\tilde T^{(l)}\) are stacked along a new mode to obtain the \emph{joint weight tensor}
\[
W\;\in\;\mathbb{C}^{\alpha^{d}\times n\times n\times L},\qquad 
W(\,\cdot\,,\,\cdot\,,\,\cdot\,,l)=\tilde T^{(l)} .
\]
The joint weight tensor is compressed with a Tucker decomposition \cite[]
{tucker1966some}, which breaks it into a small core tensor plus one factor matrix per mode, so each dimension is first projected into a lower-dimensional latent space and only the dominant interactions are kept. \cite{kossaifi2023multi} provide a detailed explanation of this construction.

\begin{figure}
    \centering
    \includegraphics[width=\linewidth]{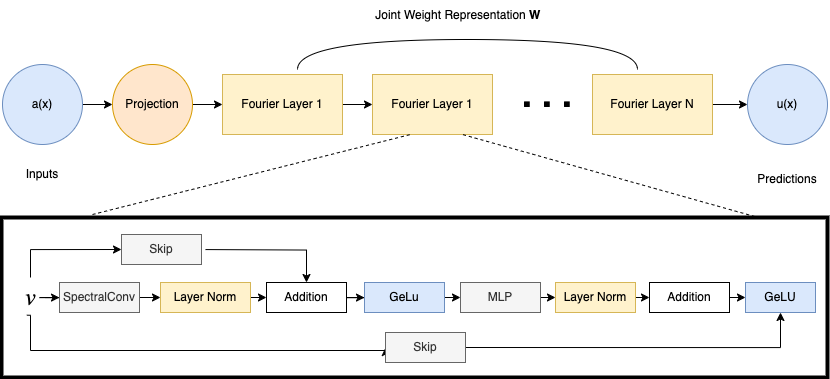}
    \caption{Architecture of the Tucker-tensorized Fourier Neural Operator. The network takes inputs $a(x)$ and lifts them to a higher-dimensional representation via a linear projection operator. The body of the network consists of $L$ Fourier layers. Each Fourier layer improves upon the base FNO architecture with layer normalization, local channel mixing, and an additional skip connection. The spectral convolution weights for all layers are gathered into a global tensor $\mathbf{W}$ and a low-rank (Tucker) decomposition is applied to reduce the number of total model parameters.}
    \label{fig:tfno_architecture}
\end{figure}

Complementing the tensor factorization, the base FNO is further improved by three orthogonal architectural refinements. First, discretization-invariant normalization is achieved by replacing Batch Normalization \cite[]{ioffe2015batch} with Layer Normalization \cite[]{ba2016layer} applied pre-activation \cite[]{he2016identity}. The normalization scale and shift parameters are shared across the entire spatial grid; this preserves mesh-independence while improving optimization stability. Second, local channel mixing is introduced by inserting a two-layer bottleneck multi-layer perceptron (MLP) after every spectral convolution, temporarily halving and then restoring the channel dimension to inject point-wise non-linear interactions that the Fourier modes alone cannot capture. Third, boundary-aware skip connections replace the original linear projections. The purpose of these linear projections is to address the limitation of traditional Fourier methods to inputs with periodic boundaries to non-periodic edges. This is critical for wave propagation problems. A gating mechanism is introduced \cite{bulat2020toward}, which smoothly interpolates between periodic and non-periodic regimes. This is complemented by a second residual link around the MLP to form a compact “double-skip” topology. Figure \ref{fig:tfno_architecture} illustrates the network architecture we use.

\subsection{Comparison of Finite Difference and TFNO}
\label{sec:fd_vs_fno}

Let $N = nx  \times ny \times nz$ denote the number of points in the domain input to the TFNO and let
\[
  N_{\mathrm{FD}} = \beta^d\,N
\]
be the number of points required by a finite-difference  scheme to satisfy dispersion and stability in \(d\) dimensions, with oversampling factor \(\beta>1\).  We write \(p\) for the finite-difference stencil width per dimension so a \((p\times p\times\cdots)\) stencil has \(\mathcal O(p^d)\) non zeros.

As discussed previously, at each time step the finite-difference scheme applies the \(p\)-wide stencil to every grid point, at cost $\mathcal O\bigl(p^d \,N_{\mathrm{FD}}\bigr) =  \mathcal O\bigl(p^d\,\beta^d\,N\bigr)$. Hence, the explicit finite-difference scheme scales asymptotically (relative to TFNO) as
\begin{equation}
     T \times \mathcal O\bigl( p^d\,\beta^d\,N\bigr),
\end{equation}
where $T$ indicates the total number of timesteps.
 
The TFNO, on the other hand, first applies a \emph{lifting} map $P: \mathbb{R}^{C_{\mathrm{in}} \times N} \to \mathbb{R}^{C_0 \times N}$ to a higher-dimensional representation, where $C_{in}$ is the number of hidden channels and $C_0$ is the number of hidden channels after lifting. Each TFNO layer then performs one forward and one inverse FFT over $N$ points with complexity $\mathcal{O}(C_0 N \log N)$ followed by dense point wise channel mixing of cost $\mathcal{O}(N C_0^{2})$ and a spectral mixing on $M$ retained Fourier modes of cost $\mathcal{O}(M C_0^{2})$. In the large-$N$ regime (e.g. 3D seismic velocities) $N>>M$ so the $M C_0^{2}$ term is lower order and the per-layer cost can be written as $\mathcal{O}\!\big(N(C_0\log N + C_0^{2})\big)$. For fixed $C_0$, this scales as $\mathcal{O}(N\log N)$, and $L$ layers therefore cost $L \times \mathcal{O}(N\log N)$ up to constant factors.

Comparing the two approaches on their respective grids for a single time step,
\[
  \underbrace{p^d\,\beta^d\,N}_{\text{FD cost}}
  \quad\text{vs.}\quad
  \underbrace{L \times N\log N}_{\text{TFNO cost}},
\]
shows that, given some value of channels, TFNO will only be asymptotically cheaper whenever
\[
  p^d\,\beta^d \;>\; L \times \, \log N.
\]

Our experimental results presented in the following sections show that even with grid coarsening and a modest layer count, the TFNO solver remains slower than the classical finite-difference methods in many configurations owing to the FFT overhead. Furthermore, observe that $p^d\,\beta^d$ grows exponentially in $d$. Consequently, if the crossover condition is \emph{not} met in three dimensions, it will not be met in two, making the 3-D timing study a conservative benchmark for any corresponding 2-D problem.

Nevertheless, asymptotic estimates alone cannot capture hardware-level realities including constant factors, memory-bandwidth pressure, and the degree of algorithmic parallelism which may scale very differently on modern GPUs. So, we treat the analysis as a guideline and rely on wall-clock benchmarks to establish the practical performance envelope.

\section*{Methods}

\subsection{Finite Difference Experiments}

The finite-difference experiments utilize reference implementations of operator stencils available in the Devito repository. For our study, we consider four classes of wave propagation, increasing in complexity and computational cost: 1) the isotropic acoustic wave equation, described in the background section, 2) the tilted-transversely isotropic acoustic wave equation, a non-physical approximation to account for anisotropic effects; we use the formulation presented by \cite{zhang2011stable}. 3) the isotropic elastic wave equation \cite[]{aki2002quantitative} and 4) the VTI elastic wave equation \cite[]{thomsen1986weak, bloot2013elastic}. We discuss the details of these various parameterizations of wave equations in the Appendix.

For each parameterization, we simulate wave propagation in a range of models of size $nx=ny=nz$ with $n(x/y/z) \in [32, 1024]$. For all experiments, the forcing term is a 10 Hz Ricker wavelet which has an upper frequency limit around 30 Hz. A 12th-order spatial discretization is used. We use layered velocity models with minimum and maximum $V_p=(1500, 3500)$ m/s and $V_s=V_p/2$. The time step ($\Delta t$) is automatically set by Devito based on an internal CFL computation, ensuring numeric stability.

We report synchronized device wall-clock time for on-GPU compute only. We instrument the interior stencil-update loops of the tiled nests and sum across tiles. This captures kernel compute cost rather than framework or allocation overhead. Memory allocation and host to device transfers are excluded here and, for parity, these are excluded in the TFNO runs.

\subsection{Fourier Neural Operator}

The Fourier Neural Operator experiments were conducted using a Tucker-tensorized FNO model (TFNO) discussed previously, with an implementation available in the \texttt{neuraloperator} repository \cite[]{kovachki2023neural}.

To generate our temporal profiling results, we assume that the network is used to simulate wave propagation in a recurrent fashion. This means timesteps are predicted sequentially (not simultaneously), i.e., $\hat u^{t+1}={G}_\theta(\hat u^{t};a)$, where ${G}_\theta$ indicates the TFNO. This is consistent with the finite difference scheme in Eq.~\ref{eq:time_stepping_general}. We make this choice because materializing the full spatiotemporal tensor $u(x,y,z,t)$ is not tractable on current GPUs without substantial engineering effort, as such tensors can reach terabyte scale. We record the elapsed time $t$ of a single forward pass and report $N_t\ \cdot \tau$ for $t$ time steps (60\, Hz Nyquist sampling gives $N_t=30$ for $\frac{1}{2}$\,s of simulation). This yields a conservative lower bound that excludes host-side loop or scheduling overhead.

Following \cite{lehmann20243d}, we also consider the ``surface-only'' wavefield configuration. In this setting we predict the wavefield $u(x,y,z{=}0,t)$, i.e. the full time history on the top surface of the 3D model. The TFNO outputs a tensor of shape $(N_x, N_y, N_t)$, where the third axis indexes time samples at the Nyquist-rate sampling interval. Unlike the recurrent rollout emulated above, this configuration produces the entire $(N_x, N_y, N_t)$ block in a single forward pass, so we report only this timing.

For the various modes of wave propagation, we assume that only the number of input and output channels will vary. The input dimensions depend on the number of physical parameters required to describe the system of equations, while the output dimensions depend on the units of the solution, e.g. pressure (1 channel) or displacement (3 channels). For isotropic acoustic simulation, we assume one input channel for velocity and one output channel for pressure. For TTI acoustic simulation, we use four input channels (Vp, epsilon, delta, theta) and two output channels. For isotropic elastic, we use three input channels (Vp, Vs, density) and three output channels associated with the displacement components of the wavefield.  Finally, for VTI elastic simulation, we use 5 input channels (Vp vertical, Vs vertical, epsilon, delta and density) and three output channels for displacement components. We find, and discuss further in the results, that the runtime does not depend meaningfully on the number of input and output channels and hence the type of parameterization.

We perform a combinatorial sweep over four key architectural knobs, number of layers \((L)\), hidden channels \((C_0)\), number of Fourier modes per dimension \((n_{\mathrm{modes}})\), and tensor rank \((r)\) on multiple cubic grids. Each unique \((L,C_0,n_{\mathrm{modes}},r,\;n_x{=}n_y{=}n_z)\) and wave-physics mode defines a \emph{configuration}. For each configuration, we conduct N=100 experiments and record the mean, standard deviation, and 95\% confidence interval of runtime. The exact value sets we use appear in Table~\ref{tab:param_grid_corr}.

During the experiments, we ran the network in its inference mode, which means that intermediate activations were not retained. This minimizes peak memory requirements. We also disabled any layers that behave differently during training (for example, dropout or normalization blocks). The model graph was statically compiled to enable on-the-fly optimization. Five warm-up passes are first conducted to ensure the compilation process has carried out all optimizations. To obtain an accurate measurement of the forward-pass evaluation time, we inserted device synchronization points immediately before and after the forward pass so that the elapsed timer reflects only the computation itself. In our implementation, these steps are carried out using \verb|PyTorch| primitives (e.g., switching between training and evaluation modes, scripting the module, and calling \verb|torch.cuda.synchronize|), but the same sequence of operations and their effects on memory, determinism, and timing apply in any modern deep-learning framework.

To compare TFNO to FD, we display measured device wall-clock runtime $t_i(V)$ as a function of volumetric size $V$. Over the tested range, both FD and TFNO are well described by linear models
\[
t_i(V)\approx \alpha_i+\beta_i V \quad (R^2>0.9).
\]
For each fixed TFNO architecture $(C_0,L)$ and for each FD physics mode, we fit a line by ordinary least squares and provide a point measure of relative runtime per configuration with the \emph{slope ratio}
\[
S=\frac{\beta_{\mathrm{FD}}}{\beta_{\mathrm{TFNO}}}.
\]
so that $S>1$ indicates TFNO is faster than FD. The slope \(\beta_i\) is the runtime per unit volume (time/volume), so the \emph{slope ratio} \(\beta_{\mathrm{FD}}/\beta_{\mathrm{TFNO}}\) is unitless and closely matches the computation-time ratio \(t_{\mathrm{FD}}(V)/t_{\mathrm{TFNO}}(V)\), but provides a point wise summary for each configuration. 

 \subsection{Compute Environment}
The simulations are run using the Lonestar6 supercomputer at the Texas Advanced Computing Center. Each simulation utilizes a single NVIDIA A100 GPU on an isolated compute node to minimize interference from background processes and eliminate contention with other users. Each node has two AMD EPYC 7763 processors with 128 cores per node and 256 GB DDR4 memory.

The full code is available at \textcolor{blue}{[URL to be released upon acceptance]}. It can be easily extended to study other PDEs of interest.

\section*{Results}

Throughout, we refer to each combination of physics mode and (for TFNO) realization of model parameters as a \emph{configuration}. Owing to our careful benchmarking design, per-configuration measurement noise was negligible for both the finite-difference and TFNO experiments. Consequently, we exclude error bars from relevant figures as they are not graphically distinguishable at plot scale. For completeness, we briefly summarize these findings.

For the FD experiments, across 34 configurations (each averaged over \(R=50\) timed runs), the median coefficient of variation (CV; standard deviation divided by sample mean) was \(0.587\%\), the 95th-percentile CV was \(2.955\%\), and the median relative 95\% confidence-interval half-width was \(0.167\%\) of the mean. Only \(4/34\) (\(11.8\%\)) configurations had relative 95\% CI half-widths \(\ge 0.5\%\); none exceeded \(5\%\), and the maximum was \(2.421\%\). 

The TFNO experimental results comprise timing measurements over all \((L,C_0,n_{\mathrm{modes}},r)\) settings (detailed in Table \ref{tab:param_grid_corr}) and physics-mode combinations. Across \(5{,}018\) configurations (each averaged over \(R=100\) timed forward passes), the median CV was \(0.44\%\), and the median relative 95\%\ CI half-width was \(0.087\%\) of the mean; \(88.4\%\) of configurations had half-widths \(<0.5\%\), the 95th percentile was \(\approx 1.10\%\), and a few outliers (relative half-width \(>5.0\%\), max \(7.61\%\)) occurred in the fastest runs (\(0.478\%\) of configurations).

\begin{figure}
    \centering
    \includegraphics[width=\textwidth]{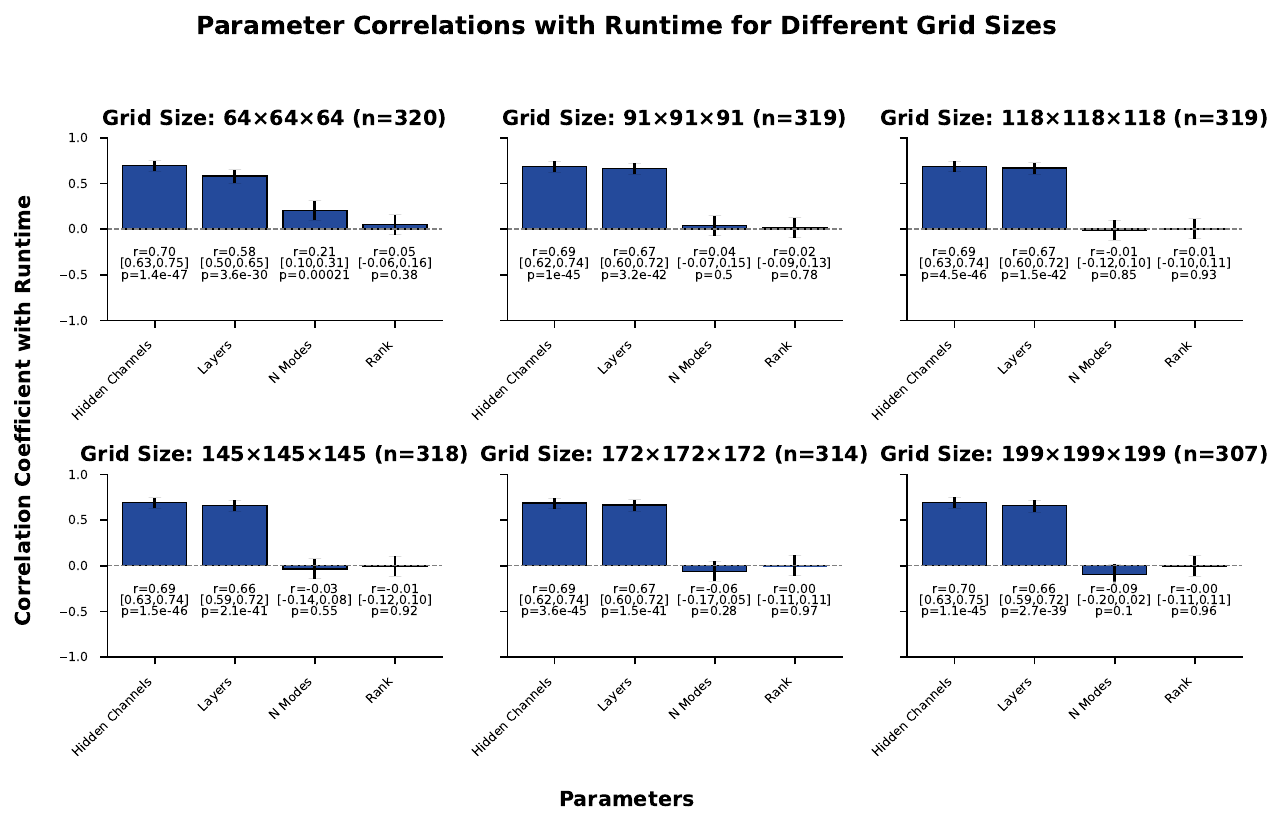}
    \caption{Pearson correlation between runtime and TFNO parameters (layers, hidden channels, number of Fourier modes, and rank) for fixed input grid sizes. Bars show the correlation coefficient $r$; error bars denote 95\% confidence interval from Fisher’s $z$ transform, and two-sided $p$-value test. Text captions report these numeric values, respectively. Subplot titles report $n$, the number of distinct parameter configurations for that grid. Runtime is the wall-clock time per forward pass, averaged over 100 measurements per configuration. Across grid sizes, hidden channels and layers exhibit the strongest positive association with runtime, whereas Fourier modes and rank have weak or negligible effects.}
    \label{fig:parameter_correlation}
\end{figure}

To understand which architectural design choices most strongly drive TFNO runtime, we compute Pearson correlations between the \emph{mean} runtime and each parameter for a subset of grid sizes \((n_x \times n_y \times n_z)\). Figure \ref{fig:parameter_correlation} shows the Pearson correlation between these network parameters and runtime as a function of the grid sizes. Figure~\ref{fig:parameter_correlation} reports Pearson $r$, 95\% confidence intervals (Fisher \(z\)), and two-sided \(p\)-values for the null hypothesis \(H_0\!:\mu=0\). Hidden channels and layers show consistent, large positive correlations \((r \approx 0.58\text{–}0.70,\; p<10^{-30})\), whereas \(n_{\mathrm{modes}}\) and \(r\) exhibit weak or no association \((|r|\le 0.21)\), with a single small but significant effect for \(n_{\mathrm{modes}}\) at the smallest grid \((64^{3})\). Consequently, we limit subsequent analysis to only these two most important network parameters. 

\begin{table}[htbp]
  \centering
  \caption{Parameter grid and timing protocol used for the correlation analysis in Fig.~\ref{fig:parameter_correlation}. For each parameter configuration we recorded the mean of $R$ timed forward passes and report a 95\% Student-$t$ CI for the mean.}
  \label{tab:param_grid_corr}

  \begin{tabular}{
    l
    l
    l
  }
    \toprule
    \textbf{Factor} & \textbf{Values tested} & \textbf{Notes} \\
    \midrule
    Layers ($L$) & $\{4,\,8,\,12\}$ & Network depth \\
    Hidden channels ($C_0$) & $\{12,\,24,\,36\}$ & Width per layer \\
    Fourier modes per dim ($n_{\mathrm{modes}}$) & $\{8,\,16,\,32\}$ & Per spatial axis \\
    Low-rank factorization rank ($r$) & $\{0.25,\,0.5,\,0.75,\,1.0\}$ & \emph{Relative} rank of tensor factorization \\
    Input grid size $G$ & $\{25^3 \,\textrm{to}\, \,253^3\}$ & One panel per $G$ \\
    \midrule
    Repeats per configuration ($R$) & $100$ & Timed forward passes \\
    Warm-up passes (discarded) ($W$) & $5$ & For graph capture and optimization \\
    \bottomrule
  \end{tabular}

  \vspace{6pt}
  \footnotesize
\end{table}

Next, we compare TFNO runtimes to Devito-generated FD code on matched spatial grid sizes $(nx=ny=nz)$ for \(0.5\) s of propagation. TFNO uses Nyquist-rate time sampling (60 Hz for a 30 Hz bandlimit), while FD advances at the CFL limit (\(\Delta t\!\approx\!1\text{–}2\) ms). As supported by the asymptotic analysis ($\mathcal{O}(n)$ for FD and $\mathcal{O}(n\log n)$ for TFNO), we did not observe TFNO to be faster than FD under any tested parameterization with the same \(10\times10\times10\) m grid spacing.

Empirically, given a fixed \((C_0,L)\) the TFNO runtime is effectively insensitive to the number of input/output channels (i.e., the wave-physics mode). After applying the lifting operator \(C_{\mathrm{in}}\!\to\!C_0\), the cost is dominated by the spectral convolutions in the dimensionality of the lifted space. Consequently, the difference in runtime between wave-physics modes is not appreciable graphically; we display all values but report the TFNO experiments under one legend item.

\begin{figure}
    \centering
    \includegraphics[width=\textwidth]{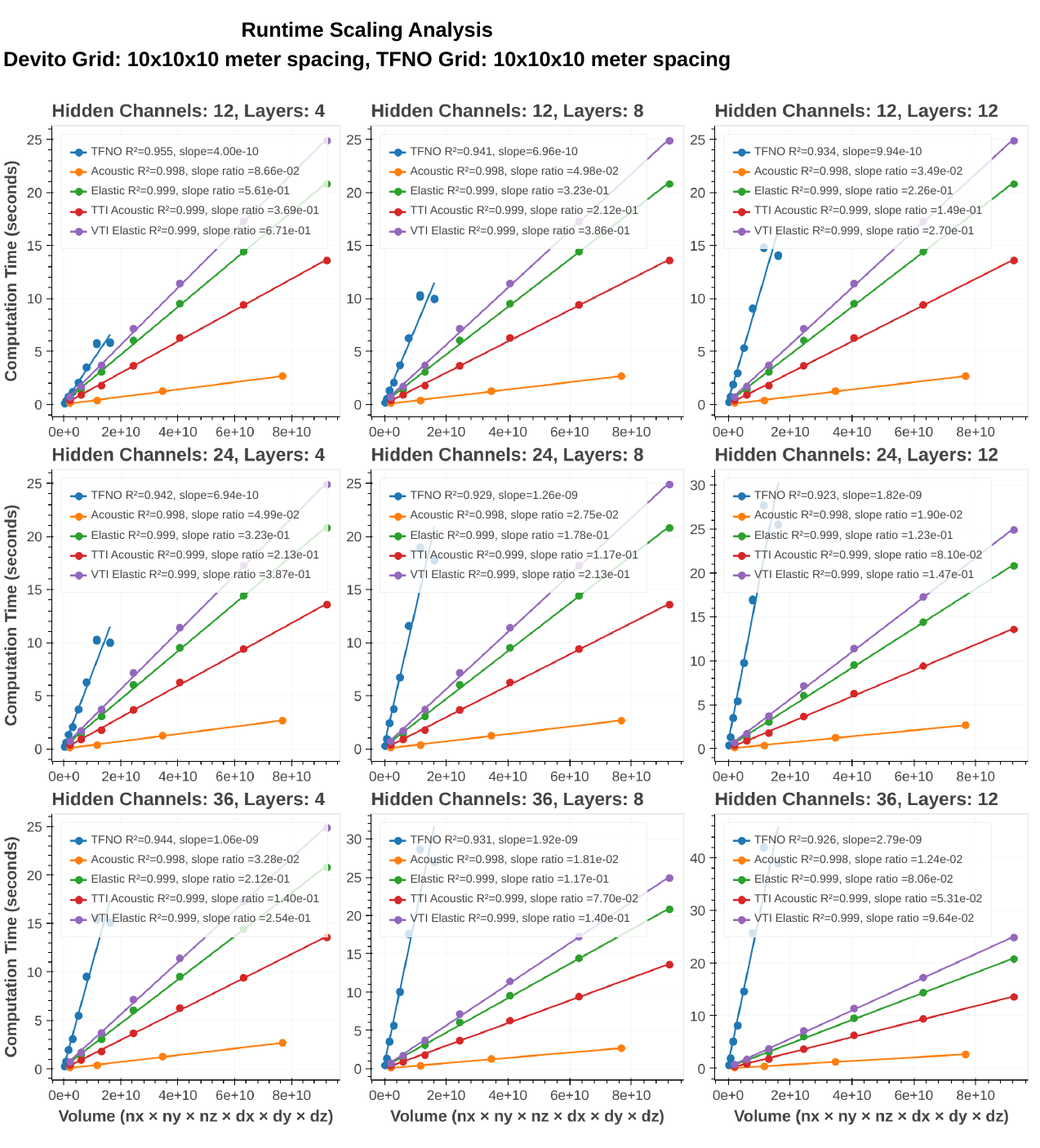}
    \caption{Runtime vs.\ volumetric extent \(V\) with matched $dx=dy=dz=10m$ grid spacing for TFNO and finite difference simulations. Linear fits achieve \(R^2>0.9\) over the sizes tested. The ratio of slopes (FD/TFNO) provides a per-configuration point measure of FD runtime relative to TFNO.}
    \label{fig:runtime_scaling_10x10x10}
\end{figure}

To compare fairly across settings we plot runtime against the \emph{volumetric size} \(V \propto n_x n_y n_z\, d_x d_y d_z\), which under uniform spacing reduces to \(V\propto n_x n_y n_z\). Because wave-propagation studies usually fix a finite physical domain rather than a grid count, this lets us compare TFNO at \(25\times25\times25\;\mathrm{m}\) spacing directly to FD at \(10\times10\times10\;\mathrm{m}\) over the same volume. Figure~\ref{fig:runtime_scaling_10x10x10} shows runtime versus \(V\) and the linear fits when the grid spacing is fixed to \(10\times10\times10\;\mathrm{m}\) for both FD and TFNO. Over the tested sizes, both methods are well described by \(t_i(V)\approx \alpha_i+\beta_i V\). We report the slope ratio, $S=\frac{\beta_{\mathrm{FD}}}{\beta_{\mathrm{TFNO}}}$, to summarize the per configuration relative runtimes.

The observed slope ratios range from \(6.71\times10^{-1}\) (reciprocating gives a \(\approx\!1.4\times\) TFNO slowdown compared to VTI elastic simulation) when \(C_0\!=\!12\), \(L\!=\!4\) up to \(1.24\times10^{-2}\) (reciprocating gives \(\approx\!80\times\) TFNO slowdown compared to acoustic) when \(C_0\!=\!36\), \(L\!=\!12\). These fits should not be extrapolated because the TFNO scales as \(N\log N\) asymptotically, but within the memory-limited regime we tested (up to \(256^3\)) the \(\log N\) factor does not induce a strong departure from linear behavior.

\begin{figure}
    \centering
    \includegraphics[width=\textwidth]{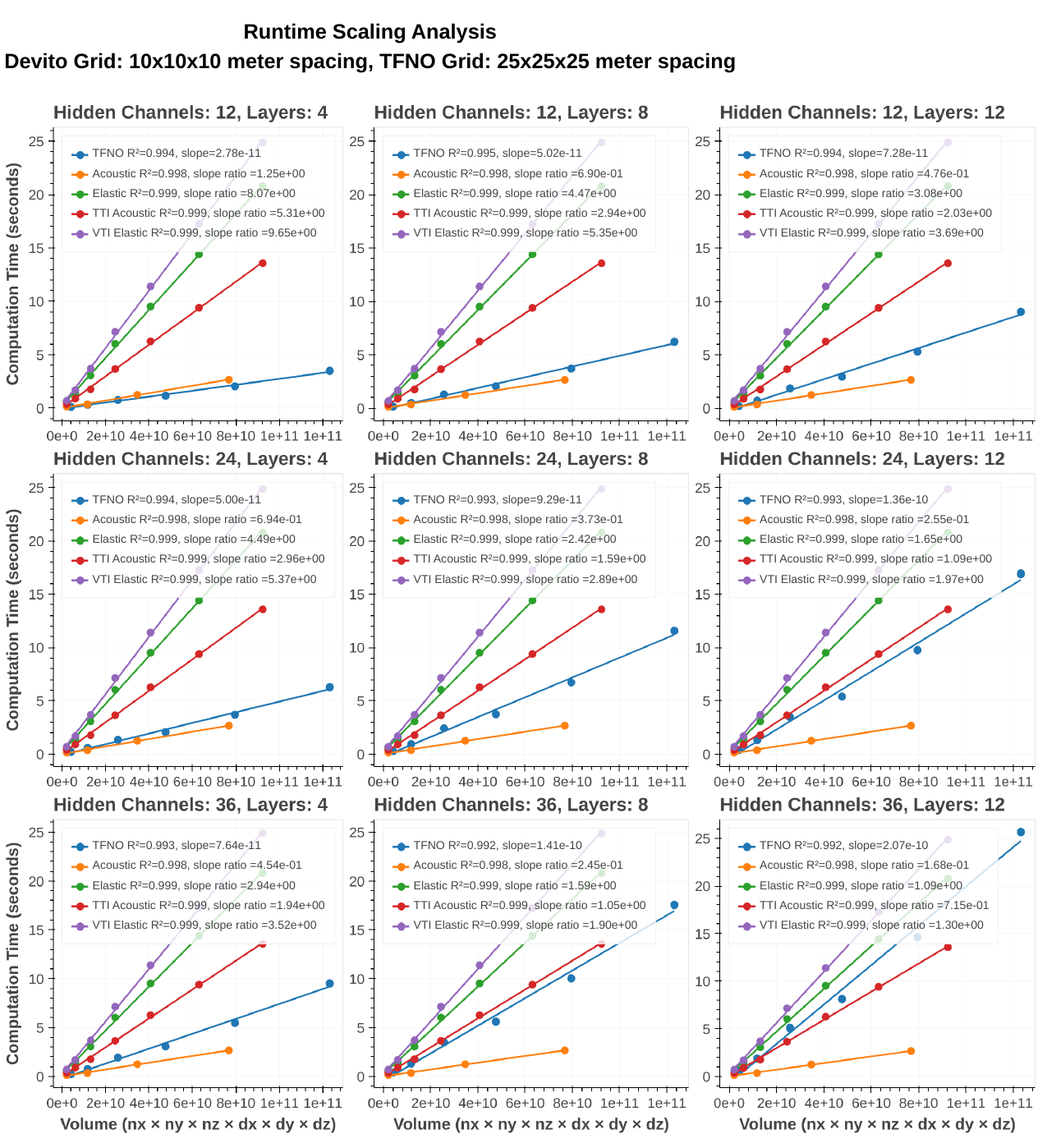}
    \caption{Runtime vs Volumetric extent. Like Figure \ref{fig:runtime_scaling_10x10x10}, but we coarsen the TFNO grid to the spatial Nyquist limit ($25 \times 25 \times 25$m) while the finite difference simulations have a $10 \times 10 \times 10$m grid spacing.}
    \label{fig:runtime_scaling_25x25x25}
\end{figure}

In the next case, the finite difference codes were run with the same 10 $\times$ 10 $\times$ 10 m grid spacing, while the TFNO is set to the equivalent physical extent with a 25 $\times$ 25 $\times$ 25 m grid spacing, leading to approximately $2.5^{3}$ fewer velocity parameters. The 25 $\times$ 25 $\times$ 25 m grid spacing is based on the spatial Nyquist frequency assuming a minimum velocity of 1500 m/s and a 30 Hz maximum frequency. We again present these results as slope ratios in Figure \ref{fig:runtime_scaling_25x25x25}.

Here, slope ratios range from $10 \times$ with $C_0=12$, $L=4$, to 1.68e-1 (reciprocating gives $\approx 6 \times$ TFNO slowdown compared to acoustic) with $C_0=36$ and $L=12$. To summarize, we tabulate a representative subset of the results based on model size in Table \ref{tab:speedup_size_physics}. 

\begin{table}[htbp]
  \centering
  \caption{Finite difference slope ratios
           \(S = \tfrac{\beta_{\mathrm{FD}}}{\beta_{\mathrm{TFNO}}}\)
           versus network size and wave-physics formulation.
           FD grid spacing \(=\;10\times10\times10\;\mathrm{m}\);
           TFNO grid spacing \(=\;25\times25\times25\;\mathrm{m}\).
           Values in {\color{red}red} indicate a slowdown (\(S<1\)).}
  \label{tab:speedup_size_physics}

  \begin{tabular}{
    l
    S[table-format=1.2]
    S[table-format=1.2]
    S[table-format=1.2]
    S[table-format=1.2]
  }
    \toprule
    & \multicolumn{4}{c}{\textbf{Speed-up (×)}}\\
    \cmidrule(lr){2-5}
    \textbf{Model size} & {Acoustic} & {TTI ac.} & {Elastic} & {VTI el.}\\
    \midrule
    TFNO-Small  &  1.25 &  5.31 &  8.07 &  9.65 \\
    TFNO-Medium & {\color{red}\num{0.37}} &  1.59 &  2.42 &  2.89 \\
    TFNO-Large  & {\color{red}\num{0.17}} & {\color{red}\num{0.71}} &  1.09 &  1.30 \\
    \bottomrule
  \end{tabular}
  \vspace{8pt}
  \footnotesize
  
  \textit{Network widths:} Small = 12 channels / 4 layers; Medium = 24 / 8; Large = 36 / 12.
\end{table}

Finally, we consider the “surface-only’’ configuration proposed by \cite{lehmann20243d}, in which we predict the full surface time history \(u(x,y,0,t)\). Here we reinterpret the third (``\(z\)'') axis as time and produce the entire block in a single forward pass (no recurrent rollout). Concretely, TFNO inputs have shape \((n_x,n_y,n_z,C_{\mathrm{in}})\) and outputs \((n_x,n_y,n_t,C_{\mathrm{out}})\) with \(n_t = n_z\) samples at a fixed, Nyquist 60\, Hz rate. For each TFNO configuration we report the elapsed time of one forward pass.

To compare with FD, we run the 3D solver for each physics class on the same \((n_x,n_y, nz)\) grid and spacing and simulate for the matching duration \(T = n_t/60\;\text{s}\). The FD step count is \(N_{\mathrm{FD}}=\textrm{ceil}({T/\Delta t_{\mathrm{CFL}}})\). Because the TFNO setup is non-recurrent, extending the time window beyond \(T\) would require additional TFNO forward passes (concatenating successive \((n_x,n_y,n_t)\) blocks).

\begin{figure}
    \centering
    \includegraphics[width=\textwidth]{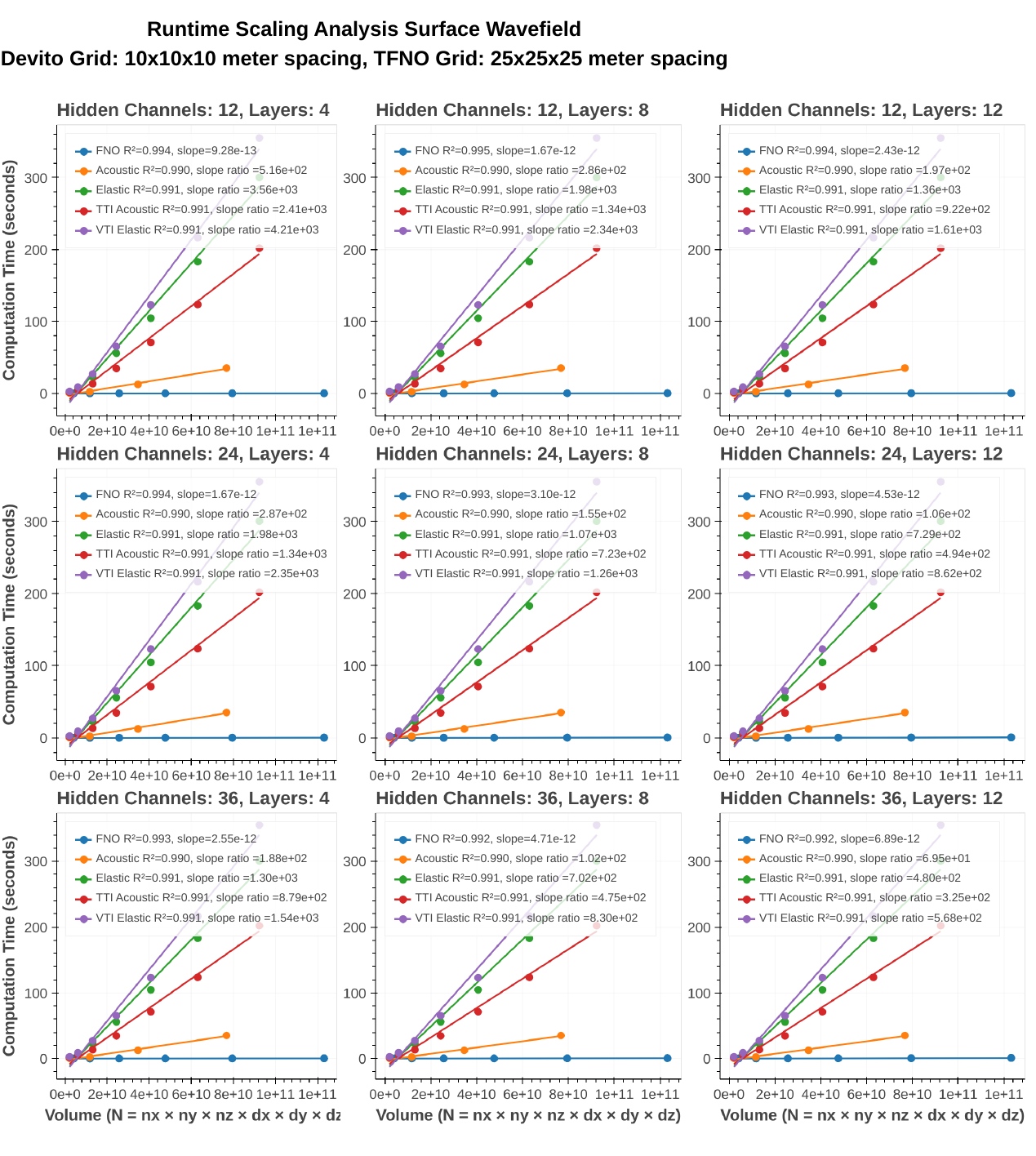}
    \caption{Runtime comparison for the surface-only wavefield \(u(x,y,z{=}0,t)\).
    TFNO produces the entire \((n_x,n_y,n_t)\) block in a single forward pass (no recurrence) by reinterpreting the third axis as time, with \(n_t=n_z\) at a fixed 60\, Hz Nyquist rate. Finite differences (Devito) run on the same spatial grid and spacing for the matching duration \(T=n_t/60\;\mathrm{s}\) using a CFL-limited time step.)}
    \label{fig:enter-label}
\end{figure}
To summarize, we tabulate these results for increasing TFNO model sizes in Table \ref{tab:surface_wavefield_speedup}. In these experiments, we observe an appreciable speedup of the TFNO, up to a factor of $4 \times 10^{3}$ relative to the finite difference simulation.

\begin{table}[htbp]
  \centering
  \caption{Finite difference slope ratios
           \(S = \tfrac{\beta_{\mathrm{FD}}}{\beta_{\mathrm{TFNO}}}\)
           when the TFNO predicts only the surface (\(z=0\)) wavefield.
           FD grid spacing = \(10\times10\times10\;\text{m}\); 
           TFNO grid spacing = \(25\times25\times25\;\text{m}\).}
  \label{tab:surface_wavefield_speedup}

  \begin{tabular}{
    l % left-aligned model names
    S[table-format=4.0] % acoustic
    S[table-format=4.0] % TTI
    S[table-format=4.0] % elastic
    S[table-format=4.0] % VTI
    }
    \toprule
    & \multicolumn{4}{c}{\textbf{Slope Ratio $S$ (×)}} \\
    \cmidrule(lr){2-5}
    \textbf{Model size} & {Acoustic} & {TTI ac.} & {Elastic} & {VTI el.} \\
    \midrule
    TFNO-Small  &  516 & 2410 & 3560 & 4210 \\
    TFNO-Medium &  155 &  732 & 1070 & 1260 \\
    TFNO-Large  &   69.5 &  325 &  480 &  568 \\
    \bottomrule
    
  \end{tabular}
  \vspace{8pt}
  \footnotesize
  
  \textit{Network widths:} Small = 12 channels / 4 layers; 
  Medium = 24 / 8; Large = 36 / 12.
\end{table}

\section*{Discussion}

The correlation analysis in Figure \ref{fig:parameter_correlation} shows that runtime varies almost exclusively with the number of layers and hidden-channel width, while Fourier-mode count and tensor-factorization rank have negligible impact.  When both TFNO and Devito were run on identical 
$10 \times 10 \times 10$ m grids (Figure \ref{fig:runtime_scaling_10x10x10}), no TFNO configuration outperformed finite differences for any of the four wave-physics formulations examined. Allowing the TFNO to operate on a coarser $25 \times 25 \times 25$ m grid, corresponding to the spatial Nyquist limit for a 30 Hz maximum frequency, produced mixed results (Figure \ref{fig:runtime_scaling_25x25x25}): the smallest network (12 channels, four layers) achieved a ten-fold speed-up for VTI elastic propagation, yet medium and large networks were between 2.7 and 6$\times$ slower than finite differences for acoustic propagation. The utility of these speedups is still contingent upon the TFNO's ability to generalize accurately at lower resolution, which itself is a significant challenge not explored here.

A more favorable regime emerges when the TFNO is tasked with predicting only the surface wavefield, $u(x,y,z=0, t)$. Here, the resulting speed-ups ranged from seventy to over four thousand, as summarized in Table \ref{tab:surface_wavefield_speedup}. These gains suggest that FNO surrogates can be highly advantageous for workflows that require repeated evaluations of boundary responses, such as marine acquisitions, where the reduction in output dimensionality and avoidance of recursive evaluation offsets the FFT overhead.

A natural extension to full-volumetric simulation that we did not consider is to treat time as an extra spatial axis and emit $K$ snapshots in a single pass. This changes (asymptotically) the per-layer cost from $\mathcal{O}\!\big(N\log N\big)$ to $\mathcal{O}\!\big(NK\log(NK)\big)$ for some choice of $C_0$. Comparing this to $K$ recurrent passes (each $\mathcal{O}(N\log N)$) gives the ratio
\[
R_K \;=\; \frac{NK\log(NK)}{K\,N\log N}
\;=\; \frac{\log(NK)}{\log N}
\;=\; 1 \;+\; \frac{\log K}{\log N}.
\]
For representative 3-D grids, $N\!\in\!\{128^3,\,192^3,\,256^3\}$ and $K\!=\!10$ yield $R_K\!\approx\!1.16,\,1.15,\,1.14$, respectively; i.e., the single-pass block is only $\sim$14–16\% slower than $K$ recurrent passes when considering the FFT term alone. In contrast, peak memory scales linearly with $K$ for the block pass (activations over $N\times K$ sites), which becomes the practical limiter as $K$ grows. Consequently, allowing a limited number of spatiotemporal snapshots does not change our compute-side conclusion relative to FD. Instead, it mainly trades host-side loop overhead for increased VRAM.

An encouraging observation across all experiments is the consistent runtime of TFNO independent of the number of input and output channels: acoustic, elastic, TTI-acoustic and VTI-elastic TFNOs exhibit virtually identical timings for a given domain size, even though they differ only in channel count. Such insensitivity is well documented in the deep-learning literature and underpins channel-wise strategies such as the multi-resolution concatenation used by \cite{phung2023wavelet}. It suggests that problems involving many distinct physical parameters or architectures that exploit channel-wise domain decomposition (e.g. \cite{kossaifi2023multi}) may benefit disproportionately from FNOs.

The gradient bottleneck in waveform-inversion remains unchanged under a TFNO surrogate: one still needs the sensitivity of a misfit functional with respect to the model. In classical FWI this is supplied by the adjoint-state method \cite[]{plessix2006review, virieux2009overview}, which forms a zero-lag cross-correlation between a forward and an adjoint wavefield, i.e., two full simulations plus checkpointing or recomputation to overcome the terabyte-scale memory footprint of 3-D problems. Reverse-mode automatic differentiation of the TFNO offers an appealingly unified route where no explicit cross-correlation is required. However, it is not cheaper: backpropagation costs at least 2 $\times$ the wall-time of a single forward pass and still stores every intermediate activation unless checkpointing is used. Moreover, the formal equivalence between reverse-mode AD and the adjoint-state gradient, proven by \cite{zhu2021general}, presumes that the network receives the full spatio-temporal wavefield; when the training data are restricted to surface traces, as in the ``surface-only" experiment, that equivalence is no longer guaranteed and the exact gradient behavior is an open research question.

Our study compared FNOs only with explicit time-domain finite-difference solvers. Spectral-element and frequency-domain finite-difference methods, both widely used in seismic modeling, may display different behavior, but evaluating those alternatives requires dedicated optimization work and is left to future work. Finally, while both Devito and TFNO codes were tuned with standard open-source optimizations and executed on identical hardware, further low-level engineering could still yield incremental speed-ups. Given the wide performance gap observed for large-capacity networks, however, such optimization is unlikely to close the one-to-two-order-of-magnitude deficit that remains between FNOs and hand-tuned finite-difference solvers in the full volumetric setting.

Overall, our results indicate that current TFNO architectures do not yet deliver net computational gains for full-volume, high-fidelity time-domain simulations particularly once training cost and inference risk are considered; even under optimistically coarse grids, larger networks remain slower than well-optimized finite-difference solvers. By contrast, for boundary-only predictions at Nyquist-rate grids, TFNOs can achieve orders-of-magnitude acceleration.

\section*{Conclusion}

A systematic wall-clock comparison between Tucker-tensorized Fourier Neural Operators and GPU-accelerated finite-difference solvers shows that the oft-cited speed-ups of FNOs largely vanish when both approaches are executed on identical hardware and equivalent spatial–temporal grids.  With a common \(10 \times 10 \times 10\;\mathrm{m}\) discretization, TFNO inference is never faster—and is frequently one to two orders of magnitude slower than GPU-accelerated finite difference across isotropic acoustic, TTI-acoustic, isotropic elastic and VTI-elastic wave equations.  Relaxing the TFNO grid to the spatial and temporal Nyquist limits yields modest gains only for very small networks (4–8 layers, 12–24 hidden channels); larger architectures incur overhead that outstrip any benefit from grid coarsening.  A decisive performance advantage emerges solely in the ``surface-only" regime, where the network predicts the $u(x,y,z=0,t)$ wavefield, delivering speed-ups of up to \(4\times10^{3}\). These findings imply that current TFNOs are not yet suitable as general-purpose surrogates for three-dimensional, time-domain, full-volume seismic modeling; however, they can be valuable accelerators for boundary-field or receiver-only applications, provided their training generalizes reliably.

\section{ACKNOWLEDGMENTS}

\append{Wave Equations}
\label{wave_equations}

We consider four classes of wave propagation, which progressively increase in complexity toward a fully anisotropic elastic wave equation.

The simplest is the mono-parameter acoustic wave equation described in the body of the paper:

\begin{equation}
    \frac{\partial^2 u}{\partial t^2} = c^2(x,y,z) \left( \frac{\partial^2 u}{\partial x^2} + \frac{\partial^2 u}{\partial y^2} + \frac{\partial^2 u}{\partial z^2} \right),
    \label{eq:wave_equation_app}
\end{equation}
where \( u = u(x, y, z, t) \) is the acoustic pressure, \( c \) is the wavespeed in the medium, and \( t \) is time.

For TTI acoustic wave propagation, we use a self-adjoint formulation \cite[]{zhang2011stable, louboutin2018effects}.

\begin{equation}
    \frac{1}{v_0^2} \frac{\partial^2}{\partial t^2} \begin{pmatrix} p \\ r \end{pmatrix} = \begin{pmatrix} G_{zz} & \sqrt{1+2\delta}(G_{xx}+G_{yy}) \\ \sqrt{1+2\delta} G_{zz} & (1+2\epsilon)(G_{xx}+G_{yy}) \end{pmatrix} \begin{pmatrix} p \\ r \end{pmatrix},
\end{equation}
where $v_0(x,y,z)$ is the velocity along the symmetry axis, $\epsilon(x,y,z)$ and $\delta(x,y,z)$ are the Thomsen parameters, $p(x,y,z)$ and $r(x,y,z)$ are stress components, and $G_{xx}$, $G_{yy}$, and $G_{zz}$ are self-adjoint rotated differential operators:
\begin{equation}
G_{xx} = (D_x)^T (D_x),
\end{equation}
where
\begin{equation}
D_x = \frac{\partial}{\partial x} \cos \phi \cos \theta + \frac{\partial}{\partial y} \sin \phi \cos \theta - \frac{\partial}{\partial z} \sin \theta.
\end{equation}

For isotropic elastic wave propagation, we use the velocity-stress formulation

\begin{align}
% Particle velocity equation
\rho \frac{\partial \mathbf{v}}{\partial t} &= \nabla \cdot \boldsymbol{\sigma} + \mathbf{f} \\
% Stress equation
\frac{\partial \boldsymbol{\sigma}}{\partial t} &= \lambda (\nabla \cdot \mathbf{v}) \mathbf{I} + \mu (\nabla \mathbf{v} + (\nabla \mathbf{v})^T),
\end{align}
where $\mathbf{v}$ is the particle velocity vector, $\boldsymbol{\sigma}$ the stress tensor, $\rho$ the medium density, $\lambda$ and $\mu$ the Lamé parameters, $\mathbf{f}$ the body force vector, and $\mathbf{I}$ the identity tensor. 

Finally,

\begin{align}
\rho \frac{\partial \mathbf{v}}{\partial t} &= \nabla \cdot \boldsymbol{\sigma} + \mathbf{f} \\
\frac{\partial \boldsymbol{\sigma}}{\partial t} &= \mathbf{C} : \boldsymbol{\varepsilon},
\end{align}
where $\mathbf{v}$ is the particle velocity vector, $\boldsymbol{\sigma}$ the stress tensor, $\rho$ the density, $\mathbf{f}$ the source term, $\mathbf{C}$ the VTI stiffness tensor with components:

\begin{align}
\mathbf{C} = \begin{pmatrix}
C_{11} & C_{11}-2C_{66} & C_{13} & 0 & 0 & 0 \\
C_{11}-2C_{66} & C_{11} & C_{13} & 0 & 0 & 0 \\
C_{13} & C_{13} & C_{33} & 0 & 0 & 0 \\
0 & 0 & 0 & C_{55} & 0 & 0 \\
0 & 0 & 0 & 0 & C_{55} & 0 \\
0 & 0 & 0 & 0 & 0 & C_{66}
\end{pmatrix}
\end{align}

and $\boldsymbol{\varepsilon} = \frac{1}{2}(\nabla\mathbf{v} + (\nabla\mathbf{v})^T)$ the strain tensor.

\newpage

\bibliographystyle{seg}  % style file is seg.bst
\bibliography{example}

\end{document}